\documentclass[comment, prd,twocolumn,nofootinbib, preprintnumbers, superscriptaddress]{revtex4}
\usepackage{amsmath,amssymb,bm,slashed,braket}
\usepackage{graphicx}
\usepackage{epstopdf}
\usepackage{float}
\usepackage{xcolor}
\usepackage{multirow}
\usepackage{dsfont}
\usepackage[colorlinks=true,
            linkcolor=blue,
            urlcolor=blue,
            citecolor=purple,          
            bookmarks=true,
            bookmarksnumbered=true,
            breaklinks=true,
            pdfpagemode=Fullscreen,
            pdfstartview=FitBH]{hyperref}
\usepackage[normalem]{ulem}
\allowdisplaybreaks[4]
\usepackage[capitalise]{cleveref}
\usepackage{orcidlink}
\usepackage{microtype}

\newcommand{\calE}{{\cal E}}

\newcommand{\calJ}{{\cal J}}

\newcommand{\calP}{{\cal P}}

\newcommand{\calS}{{\cal S}}

\newcommand{\calZ}{{\cal Z}}

\newcommand{\Avec}{{\bf A}}

\newcommand{\Evec}{{\bf E}}

\newcommand{\kvec}{{\bf k}}

\newcommand{\rvec}{{\bf r}}
\newcommand{\Rvec}{{\bf R}}

\newcommand{\Svec}{{\bf S}}

\newcommand{\CRA}{\textrm{CRA}}
\newcommand{\cco}{\textrm{cco}}
\newcommand{\ccco}{\textrm{ccco}}

\newcommand{\eV}{\textrm{eV}}

\newcommand{\nm}{\textrm{nm}}

\begin{document}

\title{Three-dimensional trapping of circular Rydberg atoms by a superimposed vortex light beam}
\author{Yi Liao\,\orcidlink{0000-0002-1009-5483}
}
\email{liaoy@m.scnu.edu.cn}
\author{Hao-Lin Wang\,\orcidlink{0000-0002-2803-5657}
}
\email{whaolin@m.scnu.edu.cn}
\author{Pengcheng Zhao\,\orcidlink{0000-0001-9211-6016}}
\email{20240264@m.scnu.edu.cn}
\affiliation{State Key Laboratory of Nuclear Physics and
	Technology, Institute of Quantum Matter, South China Normal
	University, Guangzhou 510006, China}
\affiliation{Guangdong Basic Research Center of Excellence for
	Structure and Fundamental Interactions of Matter, Guangdong
	Provincial Key Laboratory of Nuclear Science, Guangzhou
	510006, China}

\begin{abstract}
We propose to trap circular Rydberg atoms (CRAs) by a ponderomotive potential well formed with a superimposed vortex light beam. We calculate analytically the ponderomotive potential energy for a Bessel vortex light beam. We work out a corrected version of the classical circular orbit approximation for a CRA which fits the exact result much better than the usual approximation. We reveal the three-dimensional characteristics of the potential well for some benchmark values of the CRA principal quantum number and beam parameters such as the frequency, the opening angle and topological charge of the vortex. We investigate how we can achieve similar trapping effects for different principal quantum numbers by varying beam parameters. The potential provides a lattice structure in the beam axis where one CRA could be trapped at each lattice site. 
\end{abstract}

\maketitle 

%%%%%%%%%%%%%%%
\section{Introduction}
\label{sec:intro}
%%%%%%%%%%%%%%%

Optical trapping has become a daily technique in atomic physics. It is the starting point for the precise localization and manipulation of atoms with broad applications in multiple domains~\cite{marago2013optical, bradac2018nanoscale}. Rydberg atoms are highly excited atoms whose one or more electrons are in a state of very large principal quantum number. This offers them special properties such as large size, long lifetime, strong dipole interactions, and extreme sensitivity to external fields that are valuable in quantum science. For example, the strong, tunable dipole-dipole interactions between Rydberg atoms make optically trapped Rydberg systems an ideal platform for many-body quantum simulations~\cite{Browaeys:2020kzz}; and in the optical trap, high-fidelity two-qubit gates have been achieved between the ground-state and Rydberg states, which is beneficial for advancing quantum computing experiments~\cite{Bluvstein:2023zmt}.
Recently, Rydberg atoms have also been proposed as a platform for detecting light dark matter, owing to their large dipole transition moments~\cite{Engelhardt:2023qjf,Graham:2023sow,Liu:2024uwg,Chigusa:2025rqs}. Although most studies have focused on low orbital quantum number Rydberg atoms, circular Rydberg atoms (CRAs) offer superior properties for quantum simulation and computation. These are the Rydberg atoms with the largest magnitude of orbital ($l$) and magnetic ($m_l$) quantum numbers for a given principal quantum number ($n$). Their exceptionally long lifetimes significantly extend coherent interaction times compared to low-$l$ Rydberg states~\cite{nguyen2018towards} and thus lengthen the simulation time; and the enhanced coupling between neighboring CRAs helps reduce quantum gate error rates~\cite{Cohen:2021axm}. Therefore, it is important to investigate various CRA trapping methods, considering that dipole-dipole interactions between CRAs have recently been experimentally observed~\cite{Mehaignerie:2024rzo}.

Atoms may be trapped in various electromagnetic fields. Conventionally, electrostatic and magnetostatic fields~\cite{hogan2008demonstration,anderson2013production} are employed invoking large electromagnetic dipole moments of Rydberg atoms. Atoms can also be confined at intensity minima or maxima and in intermediate zones by attractive gradient forces~\cite{zhang2024full}. These trapping mechanisms originate from the linear field interaction term with charged particles in the interaction Hamiltonian. Their main drawbacks are large field-induced atomic levels shifts and radiation, thus restricting their applications in quantum computing and simulation. The ponderomotive potential, governed by the quadratic field term in the Hamiltonian, provides a more advanced trapping mechanism~\cite{dutta2000ponderomotive}. 
Experimental demonstrations have shown efficient trapping of Rydberg atoms in light field intensity minima using ponderomotive potentials, with significantly smaller field-induced levels shifts~\cite{anderson2011trapping}, which is quite helpful for studying trapped Rydberg atoms~\cite{cardman2021photoionization}. 
Furthermore, optical trapping of Rydberg atoms in ponderomotive potentials of optical bottle beams has achieved three-dimensional confinement with negligible field-induced Rydberg radiation~\cite{barredo2020three,Ravon:2023igc}. 
A so-called threading mechanism has recently been suggested~\cite{Corti_as_2024} that can simultaneously trap highly excited Rydberg states and ground-state atoms by a single tightly focused Gaussian beam.
Although blackbody-induced radiation remains a significant lifetime limiting factor~\cite{zhelyazkova2019fluorescence,cortinas2020laser}, suppression methods have recently been successfully implemented for optically trapped CRAs~\cite{Holzl:2024ajn}. 

Vortex light beams naturally feature intensity minima at their centers, making them an ideal platform for optical trapping of atoms. These beams possess intrinsic orbital angular momentum (OAM) and have been extensively studied since the seminal work by Allen et al~\cite{Allen:1992zz}. They have been generated by various devices such as cylindrical-lens mode converters~\cite{beijersbergen1993astigmatic}, spiral phase plates~\cite{beijersbergen1994helical}, and dislocated diffraction gratings~\cite{bazhenov1990laser}. However, the most widely applied method employs spatial light modulators (SLMs)~\cite{maurer2007tailoring,li2021generation}, which enable precise spatial modulation of light fields. Theoretical studies include optical trapping in Laguerre-Gaussian beams following the conventional optical force approach~\cite{he2009rotating}, and in pure Bessel vortex beams or their superposition with plane waves operating at both field maxima (red-detuned regime) and minima (blue-detuned regime)~\cite{ng2010theory} and with a resolution down to the nanometer scale~\cite{jia2020ultraprecise}. Experimentally, ultracold atom confinement has been achieved in Laguerre-Gaussian beams with minimal ac Stark shifts~\cite{kennedy2014confinement}, and two-dimensional trapping of CRAs realized in the ponderomotive potential of Laguerre-Gaussian beams~\cite{cortinas2020laser}. 
A characteristic of vortex beam trapping is the transfer of additional angular momentum to trapped particles due to non-zero OAM~\cite{ng2010theory}. This can be avoided by employing petal beams, which are formed by coherent interference of two co-propagating vortex beams with opposite topological charges but identical intensity distributions (resulting in zero net OAM)~\cite{he2009rotating}. The SLM-generated petal beams with sub-nanometer transverse confinement have been achieved~\cite{wozniak2016tighter}, and petal beams with broad tunable wavelength have also been realized at the free-electron laser (FEL) facility~\cite{Liu:2020wqd}.

In this work we propose to form an optical beam by superimposing four equal-intensity vortex beams with opposite topological charges and propagating in opposite directions. For simplicity, we illustrate the key features of such a beam with the simplest Bessel vortex. This configuration should be implementable using SLM or FEL, and provides a ponderomotive potential that traps CRAs at central field minima. It features zero net OAM, a one-dimensional optical lattice structure with three-dimensional trapping regions along the central axis, and single-CRA confinement per trapping region. The trapping potential depth and the confinement size can be adjusted by beam parameters.

This paper is organized as follows. In the next \cref{sec:theory} we present our analytical calculation for the ponderomotive potential energy for a neutral atom that is immersed in a superimposed Bessel vortex beam. The result generally applies to hydrogen-like alkali atoms and ions with one valence electron, and does not depend on how the light beam is polarized. At the end of the section we consider the case of CRAs for which a semiclassical approach provides a reasonable approximation. In \cref{sec:numerical} we reveal the characteristics of the trapping potential well for a few benchmark beam and atom parameters, and illustrate how to achieve similar trapping effect with varying atomic quantum numbers and light beam parameters. We finally conclude in the last section.

%%%%%%%%%%%%%%%%%%%%%%%%
\section{Theoretical calculations}
\label{sec:theory}
%%%%%%%%%%%%%%%%%%%%%%%%

In this section we calculate analytically the ponderomotive potential energy (PPE) generally for a neutral atom and particularly for a circular Rydberg atom (CRA) when it is immersed in a superimposed Bessel vortex light beam. We start with the potential energy experienced by a free electron, and then convolute it with the wavefunction of a bound electron in the atom. We use natural units with $\hbar=c=1$; the energy-length conversion factor for our purpose is $1~\nm^{-1}=197.3~\eV$, and the wavelength of light is related to its angular frequency by $\lambda\omega=2\pi$.

%%%%%%%%%%%%%%%%%%
\subsection{Potential energy for a free electron}
\label{sec:free}
%%%%%%%%%%%%%%%%%%

The PPE of an electron of charge $e$ and mass $m_e$ in an electromagnetic field $\Avec(t,\rvec)$ is, $e^2\Avec^2(t,\rvec)/(2m_e)$. For a monochromatic field of angular frequency $\omega$ and polarization $\vec\epsilon$, $\Avec(t,\rvec)=\vec\epsilon A(\rvec)\cos(\omega t)$, the electric field $\Evec=-\partial_t\Avec=\vec\epsilon\calE(\rvec) \sin(\omega t)$ with $\calE(\rvec)=A(\rvec)\omega$. For an applied field that oscillates rapidly compared with characteristic frequencies of the electron, we can average the ponderomotive energy over time, yielding $e^2\calE^2(\rvec)/(4m_e\omega^2)$. Since we will consider a Bessel vortex light beam which is not normalizable as a plane wave beam, we introduce a basic unit for the time-averaged ponderomotive energy of the electron, 
\begin{align}
	V^e_0=\frac{e^2\calE_0^2}{4m_e\omega^2}, 
	\label{eq:V0e}
\end{align}
which can be related to the intensity and power of the light beam. These quantities will be determined by the light beam that is experimentally realized, and we will thus cope with a dimensionless electromagnetic field from now on. 

The simplest vortex, the so-called Bessel vortex, is an equal-weight composition of plane waves that is modulated by an azimuthal phase in the plane transverse to the propagation direction of the vortex. For a vortex photon, its complex potential field reads in the system of cylindrical coordinates $\rvec=(r,\varphi_r,z)$, 
\begin{align}
	\nonumber
&	A^\mu_{\kappa mk_z\Lambda}(r,\varphi_r,z,t)
=Ne^{i(k_zz-\omega t)}
	\\
	&~\times\int\frac{d^2\kvec_\perp}{(2\pi)^2}a_{\kappa m}(\kvec_\perp)
	e^\mu_{k\Lambda}e^{+i\kvec_\perp\cdot\rvec_\perp},
	\label{eq:Bessel_kernel}
\end{align}
where $N$ is a normalization factor to be fixed soon and the kernel of the Bessel vortex is, 
\begin{align}
	a_{\kappa m}(\kvec_\perp)
	=(-i)^me^{im\varphi_k}
	\sqrt{\frac{2\pi}{\kappa}}\delta(|\kvec_\perp|-\kappa).
\end{align}
Here we choose the vortex to propagate in the $+z$ direction with angular frequency $\omega$ and momentum component $k_z$. The superimposed plane waves share the same magnitude $\kappa$ of transverse momenta $\kvec_\perp$ in the plane perpendicular to the propagation direction, so that they are monochromatic with $\omega^2=k_z^2+\kvec_\perp^2$. As we will see below, the quantity $m$, generally called the topological charge, characterizes the helical phase nature of the vortex wavefront. For a spinless vortex particle, it is the projection of an intrinsic OAM in its propagation direction. For a particle with spin, it is the projection of the total angular momentum (TAM) in its propagation direction when we choose the spin wavefunction of each plane-wave component to be a helicity eigenstate. For a photon of helicity $\Lambda=\pm 1$ and momentum $\kvec=\omega(\sin\alpha\cos\varphi_k,\sin\alpha\sin\varphi_k,\cos\alpha)$, the polarization vector is, 
\begin{align}
	\nonumber
	e_{k\Lambda}=
    &   \eta^{(-\Lambda)}e^{+i\Lambda\varphi_k}\sin^2\frac{\alpha}{2}
	+\eta^{(\Lambda)}e^{-i\Lambda\varphi_k}\cos^2\frac{\alpha}{2}
\\
    &    +\eta^{(z)}\frac{\Lambda}{\sqrt{2}}\sin\alpha,
\end{align}
where $\alpha,~\varphi_k$ are its polar and azimuthal angles. $\eta^{(\Lambda)}$ is the polarization vector of helicity $\Lambda$ for a photon moving in the $+z$ direction, and $\eta^{(z)}$ is an auxiliary vector. In the Coulomb gauge they are 
\begin{align}
	\eta^{(\Lambda)}=-\frac{1}{\sqrt{2}}(0,\Lambda,+i,0),\qquad
	\eta^{(z)}=(0,0,0,1).
\end{align}
The polarization vectors satisfy the standard orthonormal conditions, $e_{k\Lambda}\cdot k=0,~
e_{k\Lambda}^*\cdot e_{k\Lambda'}=-\delta_{\Lambda\Lambda'}$, for $\Lambda=\pm 1$. For the Bessel kernel in Eq.~\cref{eq:Bessel_kernel}, the polar angle $\alpha$ becomes the (half) angle of the cone over which the momenta of plane waves are uniformly distributed. 

Upon finishing the integral over the transverse momentum $\kvec_\perp$ in terms of the Bessel function of the first kind, 
\begin{align}
	J_n(x)=\frac{1}{2\pi}\int_0^{2\pi}d\varphi~
	e^{i(n\varphi-x\sin\varphi)},
\end{align}
and using for $\Lambda=\pm 1$, $i^\Lambda=-i^{-\Lambda}=i\Lambda$, the dimensionless Bessel photon field becomes 
\begin{align}
	\nonumber
	&A_{\kappa mk_z\Lambda}(r,\varphi_r,z,t)=\Lambda e^{i(k_zz-\omega t)}\Big[i\eta^{(-\Lambda)}e^{i(m+\Lambda)\varphi_r}\calJ_+ 
	\\
    &\qquad\qquad-i\eta^{(\Lambda)}e^{i(m-\Lambda)\varphi_r}\calJ_-+\frac{1}{\sqrt{2}}\eta^{(z)}e^{im\varphi_r}\calJ_0\Big],
\end{align}
where the following shortcuts are used,  
\begin{subequations}
\begin{align}
	\calJ_+&=J_{m+\Lambda}(\kappa r)\sin^2\frac{\alpha}{2},
	\\
	\calJ_-&=J_{m-\Lambda}(\kappa r)\cos^2\frac{\alpha}{2},
	\\
	\calJ_0&=J_{m}(\kappa r)\sin\alpha.
\end{align}
\end{subequations}
Note that we have extracted an overall scale of magnitude from the above field and parametrized it as $\calE_0/\omega$ in \cref{eq:V0e}, which amounts to choosing the normalization $N=\sqrt{2\pi/\kappa}$. As can be checked directly, the above field is an eigenstate of the $z$-projection $\hat{j}^z$ of TAM with eigenvalue $m$. From now on, we work with the three-vector potential field. Taking the real part gives the dimensionless physical field, 
\begin{align}
    \nonumber
	&\Re\Big[\vec A_{\kappa mk_z\Lambda}(r,\varphi_r,z,t)\Big]=
    \\
    \nonumber
    &+\frac{\Upsilon}{\sqrt{2}}%\cos(k_zz-\omega t)
    \big[\big(-s_+,\Lambda c_+,0\big)
	\calJ_+ -\big(s_-,\Lambda c_-,0\big)\calJ_- +\eta^{(z)}c_0\calJ_0\big]
	\\
    &-\frac{\Xi}{\sqrt{2}}%\sin(k_zz-\omega t)
    \big[\big(c_+,\Lambda s_+,0\big)\calJ_+
	+\big(c_-,-\Lambda s_-,0\big)\calJ_-
	+\eta^{(z)}s_0\calJ_0\big],
\end{align}
where $\Upsilon=\cos(k_zz-\omega t),~\Xi=\sin(k_zz-\omega t)$ and 
\begin{subequations}
\begin{align}
	c_\pm&=\cos[(m\pm\Lambda)\varphi_r],\quad s_\pm=\sin[(m\pm\Lambda)\varphi_r],
	\\
	c_0&=\cos(m\varphi_r),\qquad \quad s_0=\sin(m\varphi_r).
\end{align}
\end{subequations}

Since the Bessel function $J_n(x)$ is an oscillating, fast decaying function with the first maximal magnitude appearing at $x\sim |n|$, the above field will provide a potential well of width $\sim|n|/\kappa$ with $n=m\pm\Lambda,~m$ in the radial direction on the transverse plane. To achieve a similar confining pattern in the propagation direction and in the azimuthal direction on the transverse plane, we propose to superimpose the fields with opposite longitudinal momenta $\pm k_z$ and topological charges $\pm m$: 
\begin{align}
	\nonumber
	\vec A_{\kappa \bar m\bar k_z\Lambda}&=\frac{1}{2}\Re\Big[
	\vec A_{\kappa,m,k_z,\Lambda}+\vec A_{\kappa, m,-k_z,\Lambda}
	\\
	&\qquad\qquad\quad+
	\vec A_{\kappa,-m,k_z,\Lambda}+
	\vec A_{\kappa,-m,-k_z,\Lambda}\Big],
\end{align}
where spacetime dependence is suppressed for brevity. Noting $c_\pm\to c_\mp$, $s_\pm\to -s_\mp$, $c_0\to c_0$ and $s_0\to-s_0$ under $m\to -m$, and using $J_{-n}(x)=(-1)^n J_n(x)$ for integer $n$ and $(-1)^{m\pm\Lambda}=-(-1)^m$ for $|\Lambda|=1$, the above composed field can be simplified separately for even and odd $m$. But our goal is to obtain its time-averaged square, which for both even and odd $m$ can be written in a universal form, 
\begin{align}
	\overline{\Big\{\Re\Big[\vec A_{\kappa\bar m\bar k_z\Lambda}\Big]\Big\}^2}
	=\cos^2(k_zz)F_m(\kappa r,\varphi_r),
\end{align}
where 
\begin{subequations}
	\begin{align}
\nonumber
&      F_m(\kappa r,\varphi_r)=
       f^0_m(\kappa r)
       \\
       &\qquad\qquad\qquad 
       +4f^1_m(\kappa r)
       \cos^2\Big[m\Big(\varphi_r+\frac{\pi}{2}\Big)\Big],
		\label{eq:Fm}
		\\
&		f^0_m(\kappa r)=
		(1+\cos^2\alpha)\big[J_{m+\Lambda}(\kappa r)+J_{m-\Lambda}(\kappa r)\big]^2,
		\\
&		f^1_m(\kappa r)=
		\sin^2\alpha ~J^2_m(\kappa r)
        \nonumber
        \\
        &\qquad\qquad 
        -(1+\cos^2\alpha)J_{m+\Lambda}(\kappa r)J_{m-\Lambda}(\kappa r).
	\end{align}
\end{subequations}
A few comments are in order. The above result actually does not depend on $\Lambda=\pm 1$; i.e., for both polarizations of the laser, we have 
\begin{subequations}
	\begin{align}
		f_m^0(\kappa r)&=
		(1+\cos^2\alpha)\big[J_{m+1}(\kappa r)+J_{m-1}(\kappa r)\big]^2,
		\\
            \nonumber
		f_m^1(\kappa r)&=
		\sin^2\alpha ~J^2_m(\kappa r)
            \\
            &\quad-(1+\cos^2\alpha)J_{m+1}(\kappa r)J_{m-1}(\kappa r).
	\end{align}
\end{subequations}
This is a consequence of symmetrization in $m$. The potential is periodic in both the azimuthal and longitudinal directions with a period of $\pi/m$ and $\pi/k_z$ respectively. The effect of trapping in the direction of transverse radius is a combined consequence of the Bessel functions $J_{m\pm 1}(\kappa r)$ and $J_m(\kappa r)$. To summarize, the PPE for a free electron is, 
\begin{align}
	V^e(r,\varphi_r,z)=\frac{1}{4}V^e_0 F_m(\kappa r,\varphi_r)
	\cos^2(k_zz).
	\label{eq:Ve}
\end{align}
For $m=0$, we have $F_0=4f_0^1=4[\sin^2\alpha J^2_0(\kappa r)+(1+\cos^2\alpha)J_1^2(\kappa r)]$; i.e., there is no trapping in the azimuthal angle as expected. For large $m$, we have $F_m\approx 4J_m^2(\kappa r)\{
(1+\cos^2\alpha)-2\cos^2\alpha\cos^2[m(\varphi_r+\pi/2)]\}$, which approaches $4J_m^2(\kappa r)\sin^2[m(\varphi_r+\pi/2)]$ in the paraxial limit $\alpha\to 0$. Finally, we have also considered other superpositions of vortex photons, and found that they either do not provide confinement in some directions or are essentially the same as the above. 

%%%%%%%%%%%%%%%%%%%%%%%%%%
\subsection{Potential energy for a bound atomic electron}
\label{sec:bound}
%%%%%%%%%%%%%%%%%%%%%%%%%%

The electrons in an atom are much more affected by an external field than its nucleus because of their huge mass difference. Among the electrons the outmost ones are more loosely bound by the nucleus than the inner ones, and thus quiver more significantly in the fast oscillating field of a laser. The neutral atom can therefore be confined through the outmost electrons by a deliberately designed laser field. In this work we study the simplest case with one valence electron, the alkali atoms, such as Rubidium and Cesium, which are of daily experimental interest.  

We start with a description of atomic geometry. Suppose that the center-of-mass coordinate of the atom is $\Rvec$ and its single valence electron is located at $\rvec$ relative to it, so that the valence electron interacts with the laser at the coordinate $\Rvec+\rvec$. Then the atomic PPE is, 
\begin{align}
	V^a(\Rvec)=\int d^3\rvec~V^e(\rvec+\Rvec)|\Psi(\rvec)|^2,
\end{align}
where $\Psi$ is the wavefunction of the valence electron and $V^e$ is its free PPE computed in the last section. To simplify the matter, we assume that the atom is quantized in the same $z$ direction where the laser propagates. Since $V^e$ is obtained for a cylindrically symmetric laser, we employ the cylindrical coordinates, 
\begin{align}
	\Rvec=(P,\Phi,Z),~\rvec=(\rho,\phi,z).
\end{align}
Their sum is denoted in cylindrical coordinates as 
\begin{align}
	\Svec=\Rvec+\rvec=(\calS,\varphi,\calZ),
\end{align}
where $\calZ=z+Z$ is the longitudinal coordinate, and the transverse vector $\Svec_\perp=\Rvec_\perp+\rvec_\perp$ has the magnitude $|\Svec_\perp|=\calS$ and azimuthal angle $\varphi$, where 
\begin{align}
	\varphi=\arccos\big[\calS^{-1}\big(P\cos\Phi+\rho\cos\phi\big)\big].
	\label{eq:varphi}
\end{align}
Note that the quadrant of $\varphi$, i.e., adding $\pi$ to the above or not, is immaterial since it only appears in the form of $\cos^2[m(\varphi+\pi/2)]$ (see \cref{eq:Fm}). As the electron wavefunction will be given in terms of the spherical coordinates $\rvec=(r,\theta,\phi)$, we write $\rho=r\sin\theta,~z=r\cos\theta$, so that 
\begin{subequations}
	\begin{align}
		\calZ&=r\cos\theta+Z,
		\\
		\calS&=\sqrt{P^2+r^2\sin^2\theta+2Pr\sin\theta\cos(\phi-\Phi)}.
	\end{align}
\end{subequations}

For the valence electron in an alkali atom, the hydrogen-like wavefunction provides a good approximation: 
\begin{align}
	\nonumber
	\Psi_{nlm_l}(r,\theta,\phi)
	&=\bigg(\frac{2}{na}\bigg)^{3/2}
	\sqrt{\frac{(n-l-1)!}{2n[(n+l)!]^3}}
	\\
	&\qquad\times e^{-\tau/2}\,
	\tau^l\, L_{n-l-1}^{2l+1}(\tau)\,
	Y_l^{m_l}(\theta,\phi),
\end{align}
which satisfies the orthonormal condition:
\begin{align}
\nonumber
	&\int r^2drd\cos\theta d\phi~\Psi_{nlm_l}^*(r,\theta,\phi)
	\Psi_{n'l'm'_l}(r,\theta,\phi)
	\\
    &%\qquad\qquad\qquad
    =\delta_{nn'}\delta_{ll'}\delta_{m_l m_l'},
\end{align}
where $n,~l,~m_l$ are the principal, orbital, and magnetic quantum numbers, and $L^p_{q-p}$ and $Y_l^{m_l}$ are the associated Laguerre polynomial and spherical harmonic functions, respectively. To facilitate numerical analysis, we introduce dimensionless coordinate and momentum quantities: 
\begin{subequations}
	\begin{align}
		&r=\frac{na}{2}\tau;~
		Z=\frac{na}{2}\zeta,~P=\frac{na}{2}\calP;
		\\
		&k_z=\frac{2}{na}\xi_\parallel,~\kappa=\frac{2}{na}\xi_\perp,
	\end{align}
\end{subequations}
where $a=(\alpha m_e)^{-1}\approx 0.05292~\nm$ is the Bohr radius. The dimensionless variables appearing in $V^e(\calS,\varphi,\calZ)$ are 
\begin{subequations}
	\begin{align}
&		k_z\calZ=\xi_\parallel\left(\tau\cos\theta+\zeta\right),
		\\
&		\kappa\calS=\xi_\perp
		\sqrt{\calP^2+\tau^2\sin^2\theta+2\calP\tau\sin\theta\cos(\phi-\Phi)}.
	\end{align}
\end{subequations}

The atomic PPE becomes
\begin{align}
&	V^a(P,\Phi,Z)=\frac{(n-l-1)!}{2n[(n+l)!]^3}
	\int\tau^2d\tau d\cos\theta d\phi
	~e^{-\tau}\tau^{2l}
	\nonumber
	\\
&\qquad \times \left[L_{n-l-1}^{2l+1}(\tau)\right]^2
	|Y_l^{m_l}(\theta,\phi)|^2
	V^e(\calS,\varphi,\calZ).
\end{align}
Now we specialize to the case of a CRA with the largest orbital and magnetic quantum numbers for a given principal number $n$: $m_l=l=n-1$. Note that the result for $m=-(n-1)$ is the same. Using $L^{2n-1}_0(\tau)=(2n-1)!$, $Y_l^l(\theta,\phi)\propto e^{il\phi}(\sin\theta)^l$, and $V^e$ in \cref{eq:Ve} for a free electron, the PPE for a CRA with $|m_l|=l=n-1$ is, 
\begin{align}
	\label{eq:exactVCRA}
    \nonumber
	&V^\textrm{CRA}(P,\Phi,Z)=\frac{1}{4}V^e_0 
	\frac{1}{2n\left[2^{n-1}(n-1)!\right]^2}
	\frac{1}{4\pi}
    \\
    \nonumber
    &\qquad
    \times\int\tau^2d\tau d\cos\theta d\phi
	e^{-\tau}\tau^{2(n-1)}(\sin\theta)^{2(n-1)}
    \\
    &\qquad
    \times
	F_m(\kappa\calS,\varphi)\cos^2(k_z\calZ).
\end{align}

\subsection{Approximation with a corrected classical circular orbit}
\label{sec:ccco}

Before we embark on a numerical analysis, let us study the limit of a large principal number for a CRA. When $n\gg 1$, all dependence on $\theta$ in the integrand is mild except for the factor $(\sin\theta)^{2(n-1)}$ that abruptly drops to zero when $\theta$ deviates slightly from $\theta=\pi/2$. We may therefore evaluate the $\theta$ integral approximately by setting $\theta=\pi/2$ everywhere except in the mentioned factor: 
\begin{align}
	\int d\cos\theta~(\sin\theta)^{2(n-1)}f(\theta)
	\approx f(\pi/2)\frac{\sqrt{\pi}\Gamma(n)}{\Gamma(n+1/2)},
\end{align}
so that 
\begin{align}
	V^\textrm{CRA}(P,\Phi,Z)&\approx\frac{1}{4}V^e_0 
	\frac{1}{(2n)!}\frac{1}{2\pi}
	\int\tau^2d\tau d\phi~e^{-\tau}\tau^{2(n-1)}
	\nonumber
	\\
	&\quad
        \times
	F_m(\kappa\calS_0,\varphi_0)\cos^2(k_z\calZ_0),
	\label{eq:annulus}
\end{align}
where the subscript $0$ indicates the evaluation at $\theta=\pi/2$: 
\begin{subequations}
	\begin{align}\kappa\calS_0&=\xi_\perp
		\sqrt{\calP^2+\tau^2+2\calP\tau\cos(\phi-\Phi)},
		\\
		\varphi_0&=\arccos\big[\calS_0^{-1}\big(P\cos\Phi+r\cos\phi\big)\big],
		\\
		k_z\calZ_0&=\xi_\parallel\zeta=k_zZ.
	\end{align}
\end{subequations}
The above large-$n$ analysis suggests a semiclassical approximation. Since the wavefunction concentrates on a thin annulus lying on the transverse plane, it may be a good approximation to consider the electron moving in a classical circular orbit (cco) on the plane,  $x=r_0\cos\phi$, $y=r_0\sin\phi$, $z=0$, and to obtain the atomic PPE by averaging the free-electron PPE over the orbit. Here $r_0$ is a characteristic radius, which may be, for example, the most probable radius $\hat r=n^2a$, or the average radius $\bar r=(n^2+n/2)a$, with a small relative difference between the two of order $1/n$. The atomic PPE is then, 
\begin{align}
&	V^\textrm{CRA}(\Rvec)\approx\int_0^{2\pi}\frac{d\phi}{2\pi}\left. V^e(\rvec+\Rvec)\right|_\textrm{cco}
	\nonumber
	\\
&\quad =\frac{1}{4}V^e_0\cos^2(k_zZ)\int_0^{2\pi}\frac{d\phi}{2\pi}
	\left. F_m(\kappa\calS_0,\varphi_0)\right|_{\tau=\tau_0},
	\label{eq:semi}
\end{align}
where $\tau_0=r_0/(na/2)$. To see the relation with the thin annulus approximation in \cref{eq:annulus}, we notice that its $\tau$ integrand excluding the $F_m$ function is maximal by definition at the most probable radius $r_0=\hat r$ corresponding to $\hat\tau=\hat r/(na/2)$, with normalization: 
\begin{align}
	\frac{1}{(2n)!}\int\tau^2d\tau~e^{-\tau}\tau^{2(n-1)}=1.
\end{align}
Since $F_m$ depends mildly on $\tau$ compared with the wavefunction, this prompts us to approximate the above integrand as a delta function, so that \cref{eq:annulus} becomes 
\begin{align}
	\label{eq:cco}
	\nonumber
	V^\textrm{CRA}_\cco(P,\Phi,Z)&\approx
	\frac{1}{4}V^e_0\int\frac{d\phi}{2\pi}\int d\tau~\delta(\tau-\hat\tau)
	\\
	&\quad\times F_m(\kappa\calS_0,\varphi_0)\cos^2(\xi_\parallel\zeta),
\end{align}
which reproduces \cref{eq:semi} when $\tau_0$ is chosen to be $\hat\tau$. 

\begin{figure*}[t]
	\centering
	\includegraphics[width=0.4\linewidth]{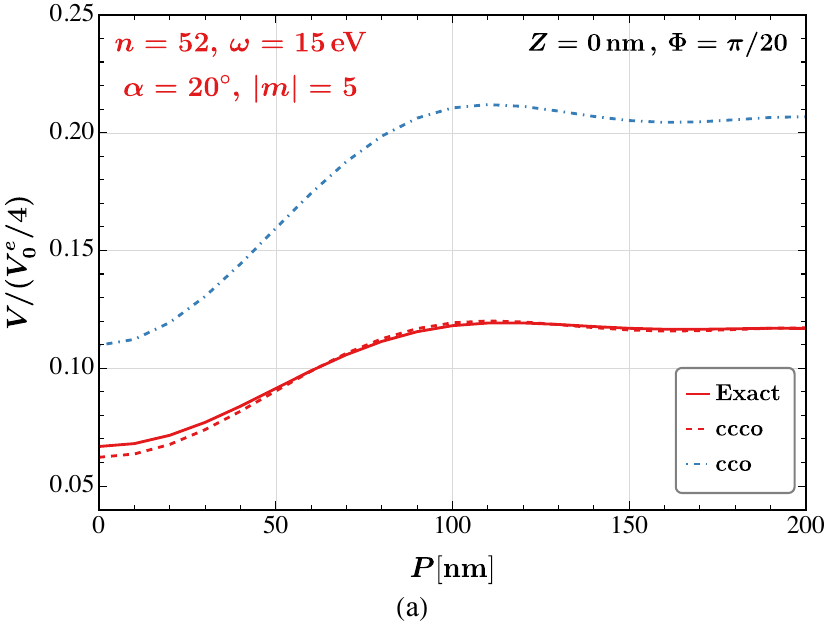}
	\qquad\qquad
	\includegraphics[width=0.4\linewidth]{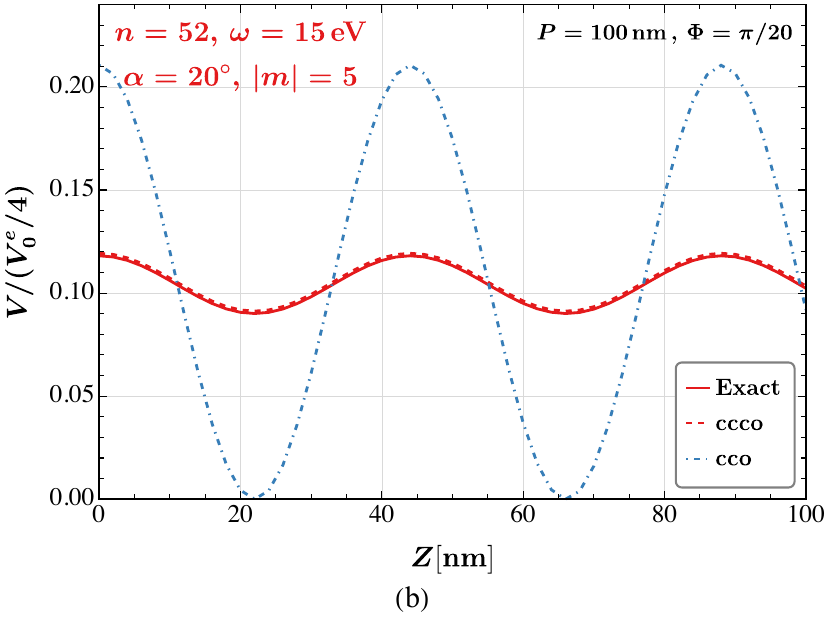}
	\caption{Comparison between the exact and cco or ccco approximate results for the atomic PPE, $V/(V_0^e/4)$, in a superimposed Bessel vortex beam. The solid (red), dashed (red), and dash-dotted (blue) curves represent the exact numerical integration, ccco and cco approximations based on \cref{eq:exactVCRA,eq:ccco,eq:cco}, respectively. The atom's position and vortex-light parameters are indicated in the figure.
	}
	\label{fig:com-approx}
\end{figure*}

However, at one point, the above argument for the semiclassical approximation may fail. Although the valence electron in a free CRA follows a classical circular orbit to a good precision, this does not necessarily mean that the approximation applies when computing the PPE for a CRA immersed in a light field. Indeed, $V^\CRA$ in \cref{eq:exactVCRA} contains a factor $\cos^2(k_z\calZ)$ that depends correlatively on the atom's position in the laser and the electron's position in the atom. In particular, it may not be a good approximation to set $\cos\theta=0$ in $k_z\calZ$ as it is multiplied by a large $\tau\sim n$. Actually, the dominant domain of $\theta$ due to the factor $(\sin\theta)^{2(n-1)}$ for $n\gg 1$ is approximately restricted to $n\cos^2\theta\ll 1$, so the magnitude of $\tau\cos\theta\sim n\cos\theta$ in $k_z\calZ$ is generally not small, and in particular not necessarily smaller than that of the longitudinal coordinate $\zeta$ of the atom. Considering this subtlety, we work out a corrected classical circular orbit (ccco) by dealing with the $\theta$ integral more properly. Noting that it is still good enough to set $\theta=\pi/2$ in the function $F_m$, the required $\theta$ integral is 
\begin{align}
	&\qquad\int d\cos\theta~(\sin^2\theta)^{n-1} \cos^2(\xi_\parallel\tau\cos\theta+\xi_\parallel\zeta)
	\nonumber
	\\
	&=\frac{\sqrt{\pi}}{2}\Gamma(n)\bigg\{\frac{1}{\Gamma(n+1/2)}
	+\frac{J_{n-1/2}(2\xi_\parallel\tau)}{(\xi_\parallel\tau)^{n-1/2}}
	\cos(2\xi_\parallel\zeta)\bigg\}.
\end{align}
The better approximate result is 
\begin{align}
	\label{eq:ccco}
	&V^\textrm{CRA}_\ccco(P,\Phi,Z)\approx\frac{1}{4}V^e_0 
	\int\frac{d\phi}{2\pi}\int d\tau~\delta(\tau-\hat\tau)
	F_m(\kappa\calS_0,\varphi_0)
	\nonumber
	\\
	&\qquad\times
	\frac{1}{2}\bigg[1 +\Gamma\bigg(n+\frac{1}{2}\bigg)
	\frac{J_{n-1/2}(2\xi_\parallel\tau)}{(\xi_\parallel\tau)^{n-1/2}}
	\cos(2\xi_\parallel\zeta)\bigg].
\end{align}
This approximation still has a simple intuitive meaning: the electron almost follows a circular orbit while making small nutation around $\theta=\pi/2$ so that the annulus is not very thin. As we will show in the next section, the approximation is very good. But if we want, we can also restore the exact $\tau$ integration: 
\begin{align}
	&V^\textrm{CRA}(P,\Phi,Z)\approx\frac{1}{4}V^e_0 
	\int\frac{d\phi}{2\pi}\int \tau^2d\tau~\frac{1}{(2n)!}e^{-\tau}\tau^{2(n-1)}
	\nonumber
	\\
	&\times F_m(\kappa\calS_0,\varphi_0)
	\frac{1}{2}\bigg[1 +\Gamma\bigg(n+\frac{1}{2}\bigg)
	\frac{J_{n-1/2}(2\xi_\parallel\tau)}{(\xi_\parallel\tau)^{n-1/2}}
	\cos(2\xi_\parallel\zeta)\bigg],
\end{align}
where the only approximation made is small nutation, but there is no simple interpretation in terms of classical orbits. 

\begin{figure*}[t]
	\centering
	\includegraphics[width=0.95\linewidth]{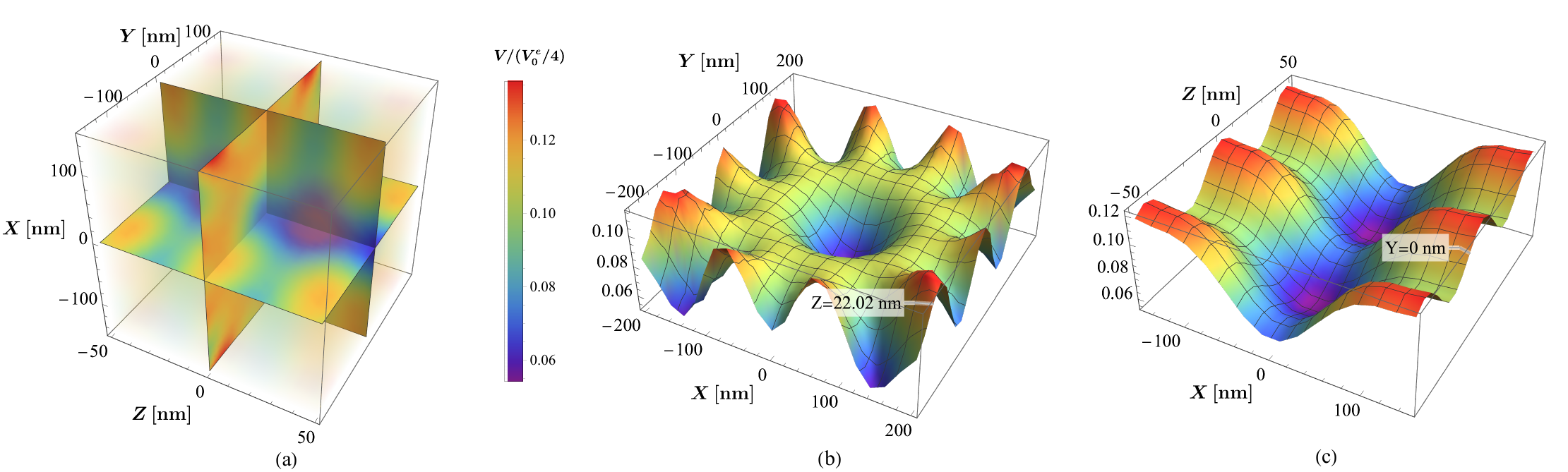}
	\caption{Three-dimensional (panel a) and two-dimensional (panels b and c) distributions of atomic PPE, $V/(V_0^e/4)$, for an $n=52$ CRA in a superimposed Bessel vortex beam with $|m|=5$, $\omega = 15$ eV ($\lambda = 82.65$ nm), and $\alpha = 20^\circ$, computed using the exact numerical integration based on \cref{eq:exactVCRA}. 
	}
	\label{fig:pp-atom-n52-m5}
\end{figure*}
%

%%%%%%%%%%%%%%%%%%%%%%%%
\section{Numerical results}
\label{sec:numerical}
%%%%%%%%%%%%%%%%%%%%%%%%

%
\begin{figure*}[t]
	\centering
	\includegraphics[width=0.95\linewidth]{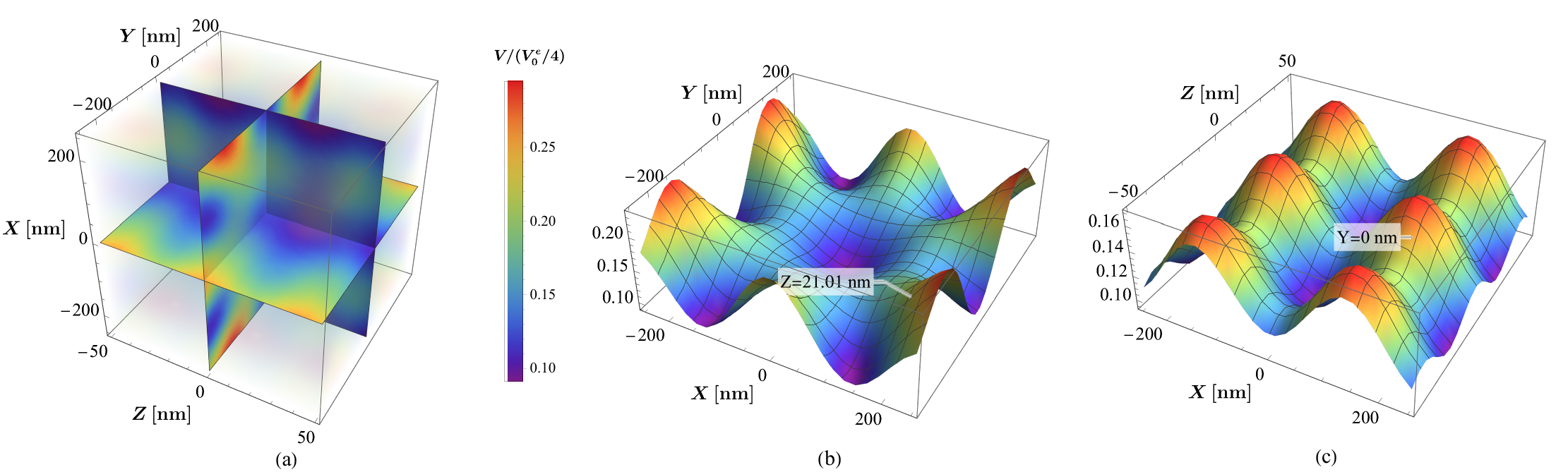}
	\caption{Similar to \cref{fig:pp-atom-n52-m5} but with $|m|=3$ and $\alpha = 10^\circ$. 
	}
	\label{fig:pp-atom-n52-m3}
\end{figure*}

\begin{figure*}[t]
	\centering
	\includegraphics[width=0.4\linewidth]{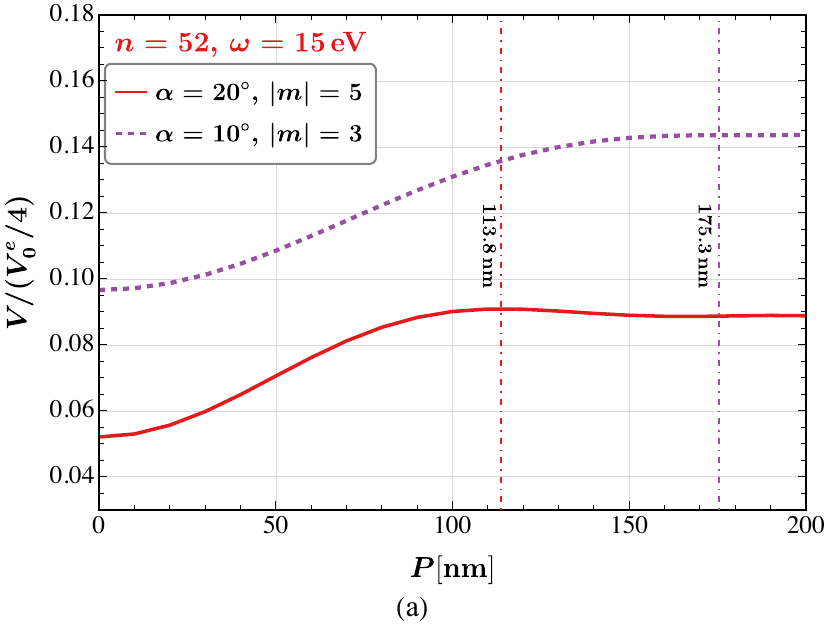}
	\qquad\qquad
	\includegraphics[width=0.4\linewidth]{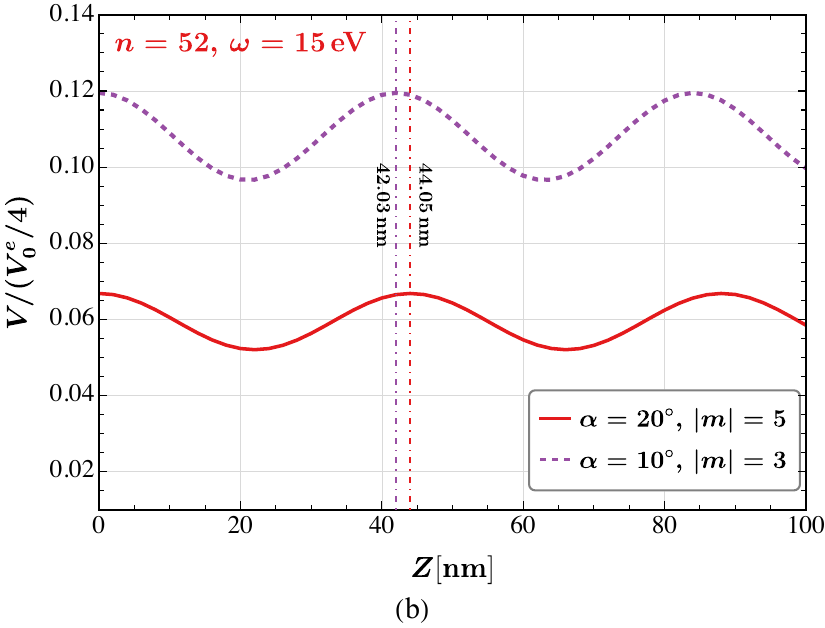}
	\caption{Atomic PPE, $V/(V_0^e/4)$, for an $n=52$ CRA is shown as a function of the transverse radius $P$ (panel a) and longitudinal coordinate $Z$ (panel b) using the same parameters as in \cref{fig:pp-atom-n52-m5} and \cref{fig:pp-atom-n52-m3}. The solid (red) and dashed (purple) curves correspond to configurations with $\alpha=20^\circ,\,|m|=5$ and $\alpha=10^\circ,\,|m|=3$, respectively. In panel (a), the longitudinal coordinate is fixed at its potential minimum: $Z=22.02$ nm for $|m|=5$ and $Z=21.01$ nm for $|m|=3$. In panel (b), the transverse radius is fixed at $P=0$.} 
	\label{fig:com-rho-Z}
\end{figure*}

\begin{figure*}[t]
	\centering
	\includegraphics[width=0.4\linewidth]{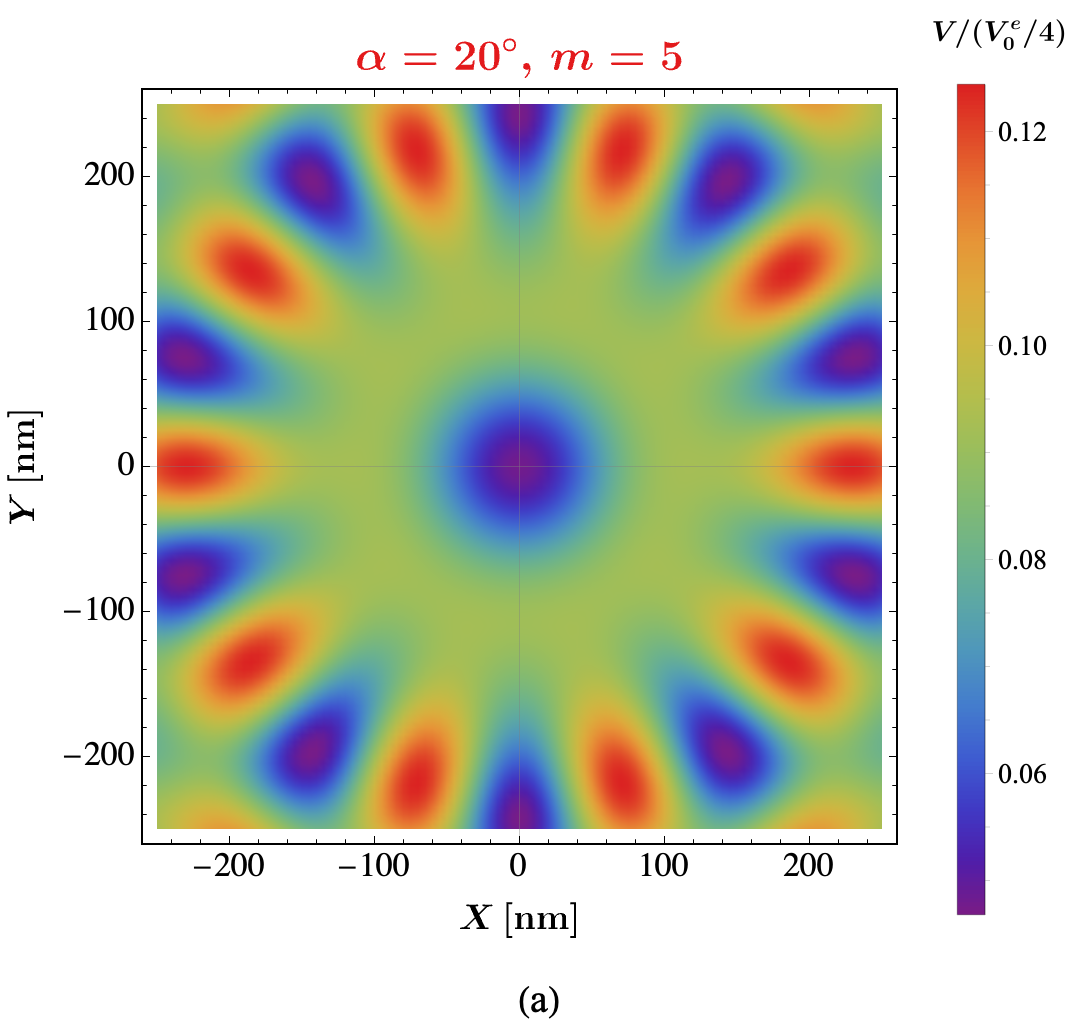}
	\qquad\qquad
	\includegraphics[width=0.4\linewidth]{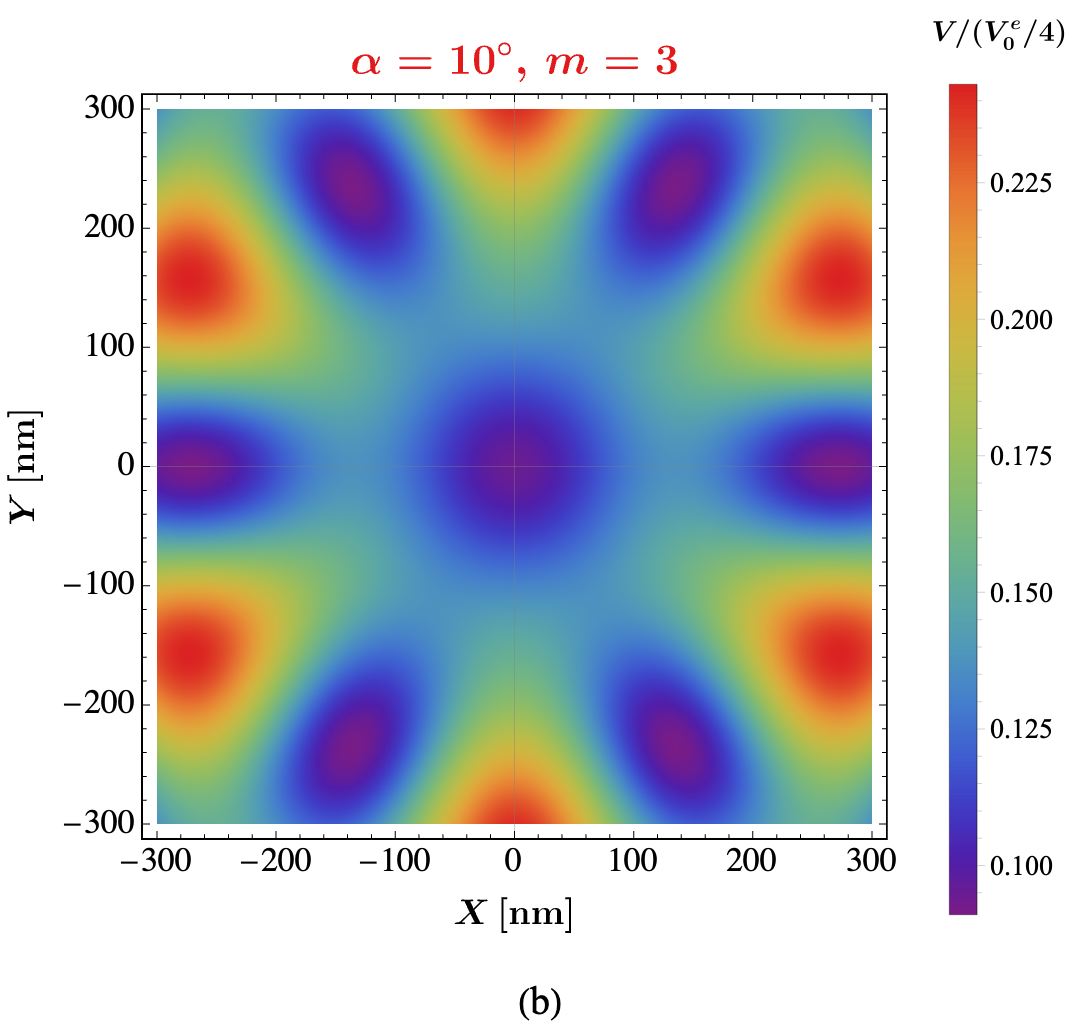}
	\caption{Petal-shaped atomic PPE in the transverse plane for (a) $\alpha=20^\circ,|m|=5$ and (b) $\alpha=10^\circ,\,|m|=3$, with the longitudinal coordinate fixed at the potential minimum, $Z=22.02$ nm and $Z=21.01$ nm, respectively.}
	\label{fig:com-trans}
\end{figure*}

\begin{figure*}[t]
	\centering
	\includegraphics[width=0.31\linewidth]{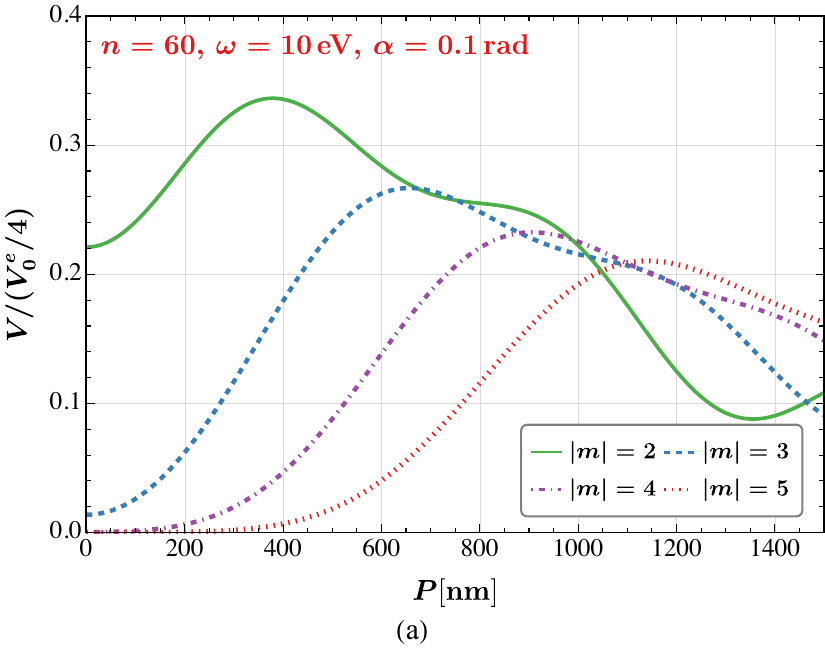}
	\quad
	\includegraphics[width=0.31\linewidth]{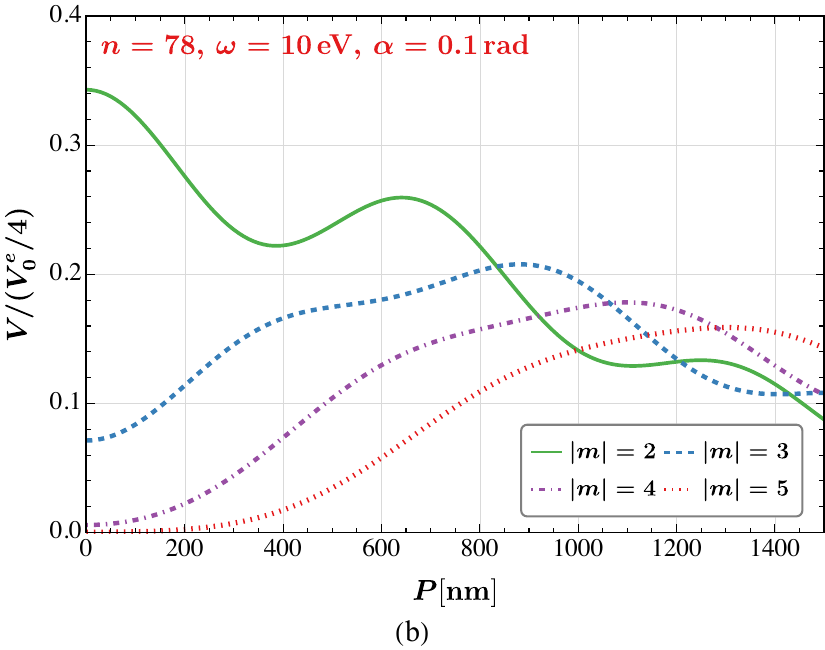}
	\quad
	\includegraphics[width=0.31\linewidth]{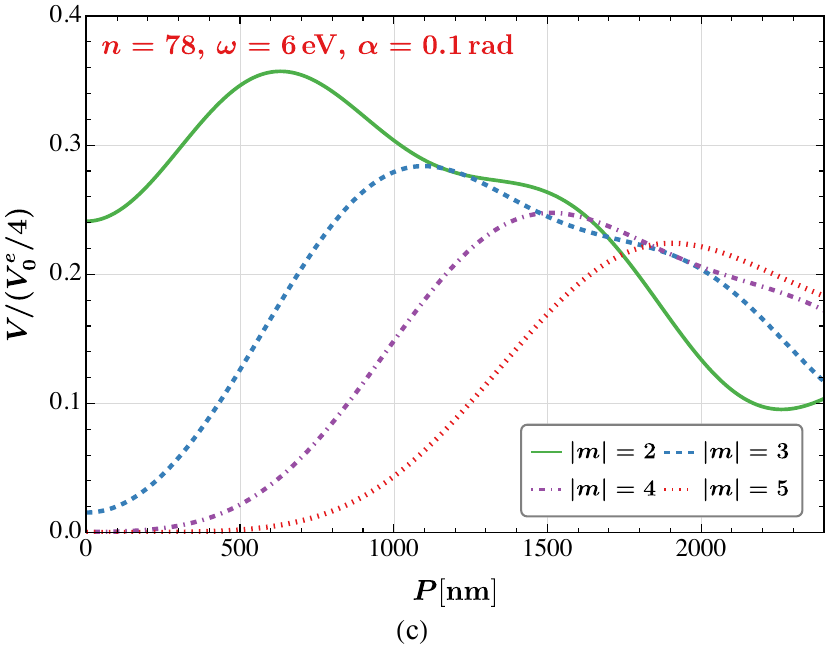}
	\caption{Atomic PPE for a CRA, demonstrating the impact of the principal quantum number $n$ and vortex light beam parameters for various values of $|m|$. 
	}
	\label{fig:diffm}
\end{figure*}

In this section we analyze the performance and characteristic features of the proposed trapping scheme. We first compare in \cref{fig:com-approx} the exact result based on \cref{eq:exactVCRA} with the two approximate results in \cref{eq:cco,eq:ccco} for some values of the atom and vortex light parameters. It can be clearly seen that our ccco approximation shows excellent agreement with the exact result, while the approximation based on the usual cco generally exhibits large discrepancies. Comparing \cref{eq:cco,eq:ccco} one sees that the two approximations cannot coincide for general coordinates of the atom since the coefficient of $\cos(2\xi_\parallel\zeta)$ in the square brackets of \cref{eq:ccco} reaches its maximal value 1 at $\xi_\parallel\tau=0$ and then drops rapidly in magnitude with increasing $\xi_\parallel\tau$. The only exception occurs at $\cos(2\xi_\parallel\zeta)=0$, i.e., where the atom is located longitudinally at $Z=(2\ell+1)\pi/(4k_z)$ with $\ell$ being an integer. This explains the crossing of the curves shown in the right panel of \cref{fig:com-approx}. 

Now we investigate the three-dimensional profiles of the atomic PPE. 
As a benchmark example, we consider an $n=52$ CRA. In \cref{fig:pp-atom-n52-m5}(a) the three-dimensional trapping pattern is shown for triple slices of planes passing through the origin, and in \cref{fig:pp-atom-n52-m5}(b) and (c) the two-dimensional trapping patterns are shown in the $XY$ and $ZX$ planes at $Z=22.01~\nm$ and $Y=0$ respectively. These patterns arise from a superimposed Bessel vortex beam with topological charges $|m|=5$, photon energy $\omega = 15~\eV$ ($\lambda = 82.65~\nm$), and opening angle $\alpha = 20^\circ$. The color density represents the dimensionless ponderomotive potential energy, $V/(V_0^e/4)$. In this configuration, the CRA is effectively located at the center of the transverse plane and within the trough of the longitudinal potential well. Since the longitudinal component of the wave vector exceeds the transverse one ($|k_z|>\kappa$) for a practical opening angle $\alpha<45^\circ$, the potential well is broader in the transverse direction ($\sim 113.5$ nm) compared to the longitudinal axis ($\sim 22$ nm). This geometry is particularly well suited to confining the thin annular structure of a CRA. 
We further analyze an $n=52$ CRA trapped in a superimposed vortex light beam with $|m|=3,~\alpha=10^\circ,~\omega = 15~\eV$, and the result is shown in \cref{fig:pp-atom-n52-m3}. Both parameter settings in \cref{fig:pp-atom-n52-m5} and \cref{fig:pp-atom-n52-m3} satisfy the condition $\kappa \hat{r} \sim |m|-1$, ensuring that the atom is confined to the center of the transverse plane by a potential barrier that is dominated by the first local maximum of the Bessel functions. The resultant PPE, $V/(V_0^e/4)$, as a function of the transverse radius $P$ and longitudinal coordinate $Z$, is displayed in \cref{fig:com-rho-Z}. Compared to the $|m|=5$ case, the $|m|=3$ configuration produces a broader potential well in the transverse radial direction ($P$), accompanied by a slightly narrower confinement along the longitudinal axis ($Z$). Modulations in the azimuthal direction are visualized in \cref{fig:com-trans}, where a petal-shaped structure of the atomic PPE is determined by the topological charge $m$. It can be seen clearly from the figure that the depth of the potential well in the azimuthal direction increases with the radius $P$, which helps stably trap the atom at the center of the transverse plane as we have discussed above.

The periodic structures in the longitudinal and azimuthal directions witnessed in \cref{fig:com-rho-Z}(b) and \cref{fig:com-trans} can be understood as follows. The $Z$ dependence in the potential $V^\CRA$ (see \cref{eq:exactVCRA}) only appears in the last factor $\cos^2[k_z(r\cos\theta+Z)]$ which is obviously periodic in $Z$ with a period of $\pi/k_z$. The $\Phi$ dependence in $V^\CRA$ enters implicitly through the angle $\varphi$ in the form $\cos^2[m(\varphi+\pi/2)]$ (see \cref{eq:Fm}). Since the integrand of the $\phi$ integral is periodic with a period of $2\pi$, one can shift it by $\Phi$, $\phi\to\phi+\Phi$, while keeping its integration domain $[0,2\pi]$ intact, so that \cref{eq:varphi} yields $\cos\varphi=\cos(\Phi+\delta)$, where $\cos\delta=(P+\rho\cos\phi)/\bar\calS$ and $\sin\delta=\rho\sin\phi/\bar\calS$ with $\bar\calS=\calS|_{\phi-\Phi\to\phi}$ are independent of $\Phi$. Thus, $\varphi=\pm(\Phi+\delta)\textrm{ mod }2\pi$, and the periodicity of $V^\CRA$ in $\Phi$ now becomes evident, and the period is $\pi/m$. 
The periodic structure in the propagation direction of light provides a lattice trapping scheme in which one CRA is trapped at each lattice site in the beam axis.

Finally we assess the effects on trapping a CRA of a different principal quantum number $n$ by varying parameters of a vortex light beam. The topological charge $m$, which governs the helical phase structure of vortex light, plays a crucial role in shaping the potential well. In \cref{fig:diffm} (a), we show the PPE for an $n=60$ CRA with the beam paramters $\omega = 10~\eV$ and $\alpha = 0.1\textrm{ rad}\approx 5.7^\circ$ but different topological charges. Variations in $m$ result in different changes in the width and depth of the potential well. Specifically, for $|m| = 2, 3, 4, 5$, the width of the potential well increases with $|m|$, while the depth first increases and then decreases. As the principal quantum number $n$ increases, the corresponding cco radius $\hat r$ becomes larger, shifting the trapping potential to smaller $P$ values, as shown in \cref{fig:diffm}(b). This effect can be compensated for by reducing the magnitude of the transverse momentum $\kappa$. As illustrated in \cref{fig:diffm}(c), a similar trapping effect can be achieved for an $n=78$ CRA with a lower frequency $\omega = 6$ eV or a smaller opening angle $\alpha$. For CRAs with principal quantum numbers $n_1$ and $n_2$, the required transverse momenta $\kappa_i$ should satisfy the relation $n_1^2 \kappa_1 \approx n_2^2 \kappa_2$ for proper trapping. This suggests that our trapping scheme is more advantageous for higher-$n$ CRAs with lower-frequency vortex light, particularly for a given open angle $\alpha$.

%%%%%%%%%%%%%%%%%%%%%%%%%%
\section{Conclusions}
\label{sec:conclusions}
%%%%%%%%%%%%%%%%%%%%%%%%%%

In this work, we have proposed a three-dimensional trapping scheme for circular Rydberg atoms (CRAs) using a superimposed vortex light beam. Confinement in the transverse plane is achieved with vortex light carrying opposite topological charges, $\pm m$, while longitudinal confinement arises from fields with opposite longitudinal momenta, $\pm k_z$.

We derived the ponderomotive potential energy (PPE) experienced by a free electron and convoluted it with the wavefunction of a bound electron to obtain the PPE for the atom. This analytical result is general for any neutral atom with one valence electron. We also developed a corrected classical circular orbit approximation for CRAs in the limit of large principal quantum number $n$, which was numerically verified and found to agree excellently with the exact result. The resulting three-dimensional trapping potential was explored for several benchmark parameter choices. The atomic PPE profiles demonstrate that the CRA is effectively confined at the center of the transverse plane and located within the trough of the longitudinal potential well. Importantly, the transverse confinement is broader than the longitudinal confinement, highlighting the feasibility of our scheme for trapping the thin annular structure of a CRA. Furthermore, the periodic structures in both longitudinal and azimuthal directions were numerically confirmed. The longitudinal periodicity gives rise to a lattice structure, with one CRA trapped at each lattice site along the beam axis, while the azimuthal periodicity produces a petal-like pattern in the transverse plane that enhances the stability of the central trap. Finally, we examined the effects of varying the principal quantum number $n$ on the trapping of CRAs. By adjusting the parameters of the vortex light beam, we found that for CRAs with principal quantum numbers $n_1$ and $n_2$, the transverse momenta $\kappa_i$ should satisfy the relation $n_1^2 \kappa_1 \approx n_2^2 \kappa_2$ for proper trapping. This scaling indicates that the scheme is particularly advantageous for higher-$n$ CRAs when using lower-frequency vortex light, especially for a given open angle $\alpha$.

%%%%%%%%%%%%%%%%%%%%%%%%%%
\section*{Acknowledgements}
%%%%%%%%%%%%%%%%%%%%%%%%%%
This work was supported in part by Grants 
No.\,NSFC-12035008, 
No.\,NSFC-12247151, 
and No.\,NSFC-12447117.

\bibliography{refs-draft}{}
\bibliographystyle{utphys}

\end{document}